# Energy deposition studies for the LBNE beam absorber[*]

**Igor L. Rakhno[1], Nikolai V. Mokhov[1], Igor S. Tropin[1]**
[1]Fermi National Accelerator Laboratory, Batavia, Illinois 60510, USA

**Abstract**

*Results of detailed Monte Carlo energy deposition studies performed for the LBNE absorber core and the surrounding shielding with the MARS15 code are described. The model of the entire facility, that includes a pion-production target, focusing horns, target chase, decay channel, hadron absorber system – all with corresponding radiation shielding – was developed using the recently implemented ROOT-based geometry option in the MARS15 code. This option provides substantial flexibility and automation when developing complex geometry models. Both normal operation and accidental conditions were studied. Various design options were considered, in particular the following: (i) filling the decay pipe with air or helium; (ii) the absorber mask material and shape; (iii) the beam spoiler material and size. Results of detailed thermal calculations with the ANSYS code helped to select the most viable absorber design options.*

___

## Introduction

The Long-Baseline Neutrino Experiment (LBNE) at Fermilab is supposed to provide the world's highest-intensity neutrino beam for the US program in neutrino physics [1]. The corresponding incoming proton beam power can ultimately be as high as 2.3 MW, and the underground beam absorber at the end of the decay channel with related infrastructure is supposed to operate with little or no maintenance for about 20 years. Such a combination of long operation time and high deposited power imposes strict limitations on design of the absorber. In this paper, both normal operation and accident are studied. All the calculations described below were performed with the MARS15 Monte Carlo computer code [2-3].

## Normal operation and accidents

### *Normal operation*

At normal operation, the 120-GeV proton beam delivered to the target will ultimately have $1.6 \times 10^{14}$ proton/pulse with time interval between pulses being 1.33 second. The average beam power on the target will be equal to 2.3 MW. The beam shape at this location is described by Gaussians with $\sigma_x = \sigma_y = 1.3$ mm and $\sigma_{x'} = \sigma_{y'} = 17$ μrad. The beam is tilted down by 101 mrad.

### *Accidents*

Development of credible accident scenarios usually requires separate investigations. At this stage, we studied only a 'target disappears' scenario which, in a sense, represents the most severe case in terms of power deposited in the absorber core. Accident scenarios with a misbehaved beam will be developed and studied later. Such scenarios will be useful, for example, in order to determine if cooling channels, located at some distance from the absorber core center, can withstand such accidents.

## Unified computer model

The elevation views of the entire MARS model and several fragments—target hall, target chase and absorber hall—are shown in Figures 1 thru 3. The absorber model shown in Figure 2 represents one of the most recent design options with a beam spoiler. Figure 3 shows that a lot of attention was paid to tiny details in order to follow the design specifications as close as possible. The color code used in the Figures implies that light blue and gray colors usually refer to air and concrete while meaning of the other colors depends on the problem studied.

**Figure 1: Elevation view of the entire unified model**

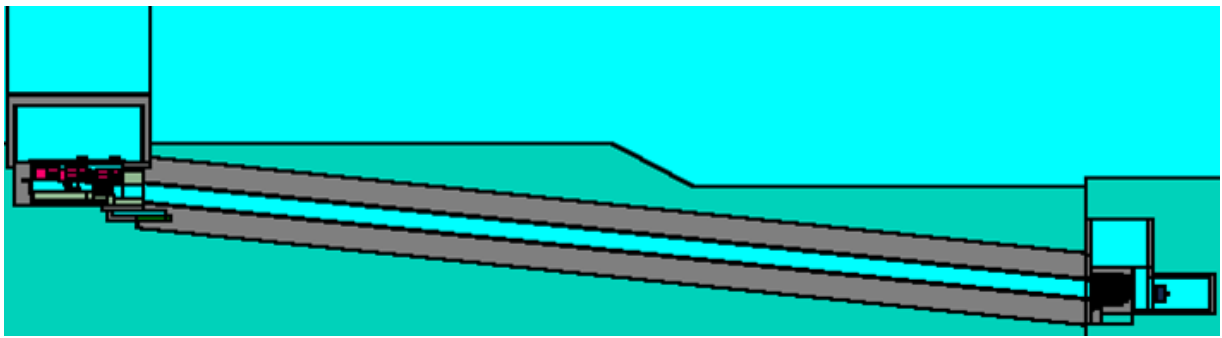

**Figure 2: Elevation view of the leftmost (target hall, top) and rightmost (absorber, bottom) fragments of the entire computer model**

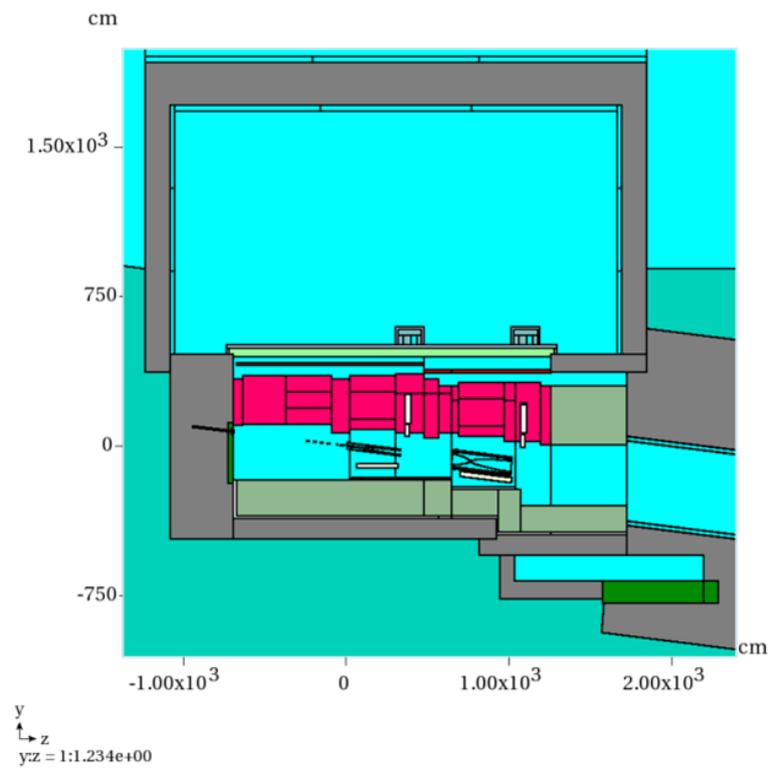

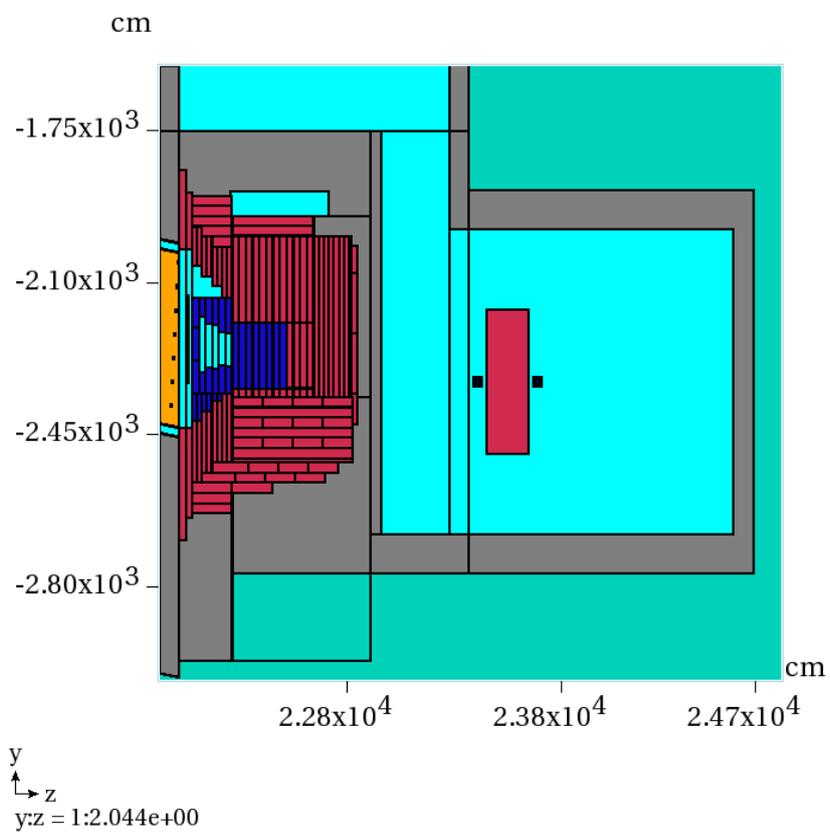

**Figure 3: Two fragments of the target chase model**

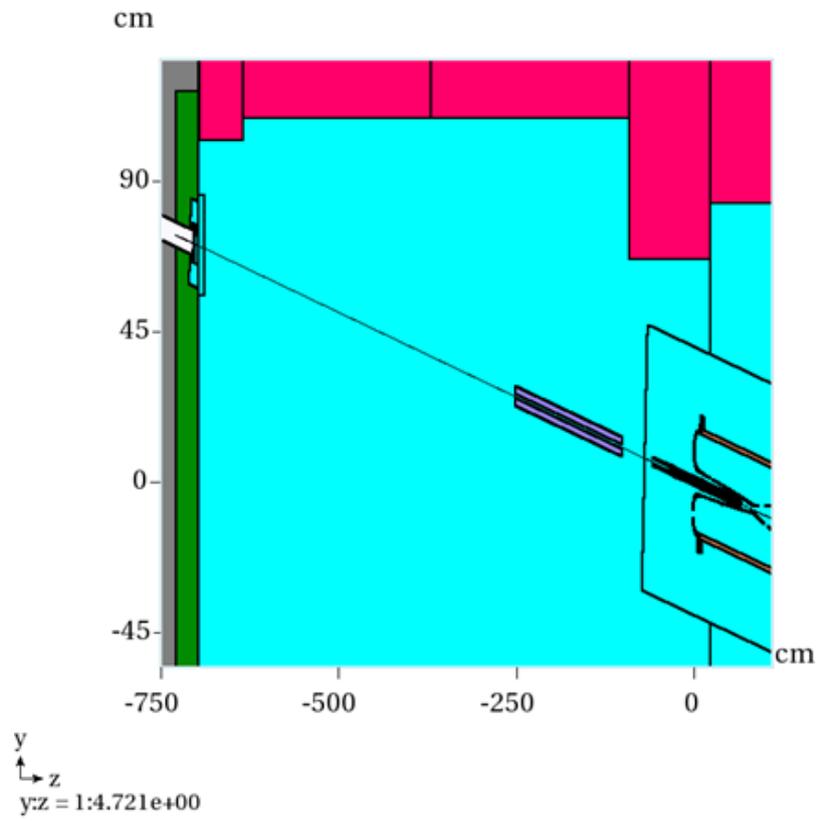

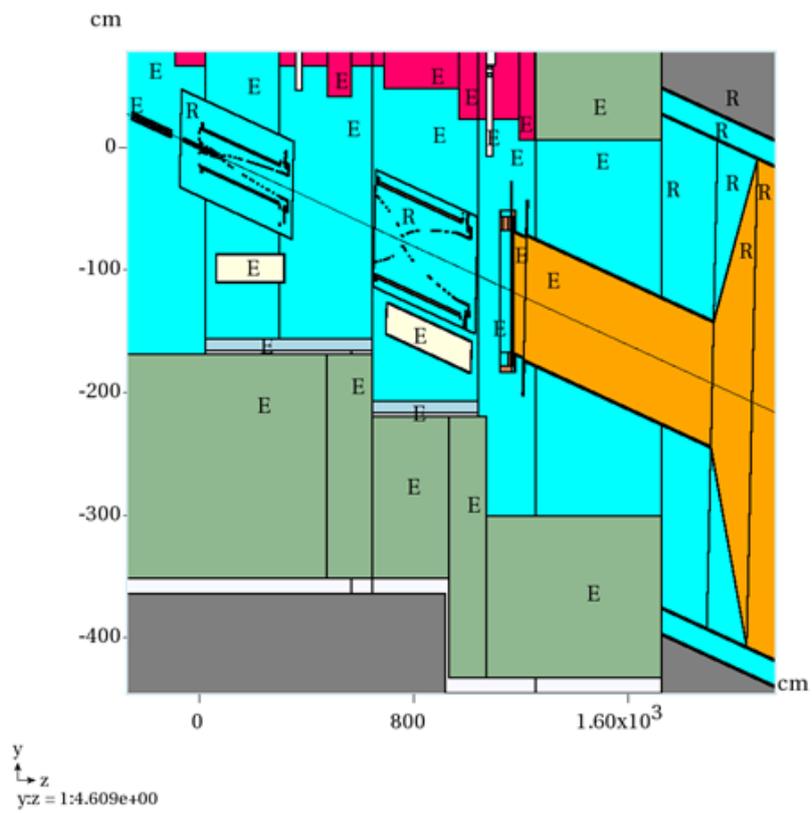

**Source term**

The incoming beam for the absorber was calculated with the MARS15 code in the exclusive mode using the LAQGSM hadron generator. Interactions in both the target and decay channel were taken into account, so that one can predict the effect due to replacement of the air with helium in the decay pipe. Four different cases were considered: normal operation and an accident for both the air and helium as a filling gas in the decay pipe. The calculated distributions of the source term across the beam pipe cross section are shown in Figures 4 and 5. One can see that the area with major energy deposition in the absorber core is expected to be confined within two feet in radius. And due to reduced scattering on helium in the decay pipe, the helium case is more severe in terms of peak deposited energy for both normal operation and accident. At the same time, the accident case allows for simplified analytical calculations for the beam window, and at the end of the decay pipe the beam can be represented with a Gaussian as shown in Figure 6.

**Power density distributions**

The calculated deposited power for normal operation with helium in the decay pipe is shown in Table 1. More detailed power density distribution for the aluminum core is shown in Table 2. The calculated deposited power density distributions for normal operation are shown in Figure 7. One can see that in the case with helium the peak power density is higher by about 30% compared to the case with air.

**Figure 4: Calculated particle and energy flux (arbitrary units) for normal operation with air (left) and helium (right) in the decay pipe**

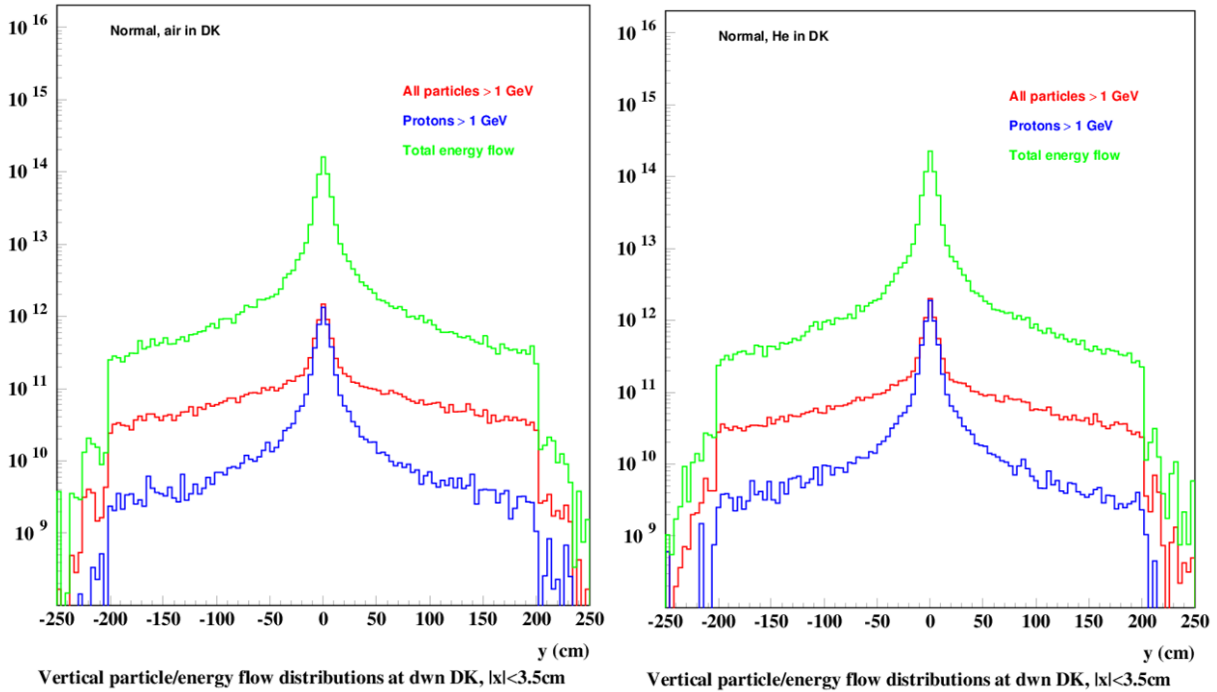

**Figure 5: Calculated particle and energy flux (arbitrary units) for the accident with air (left) and helium (right) in the decay pipe helium (right) in the decay pipe**

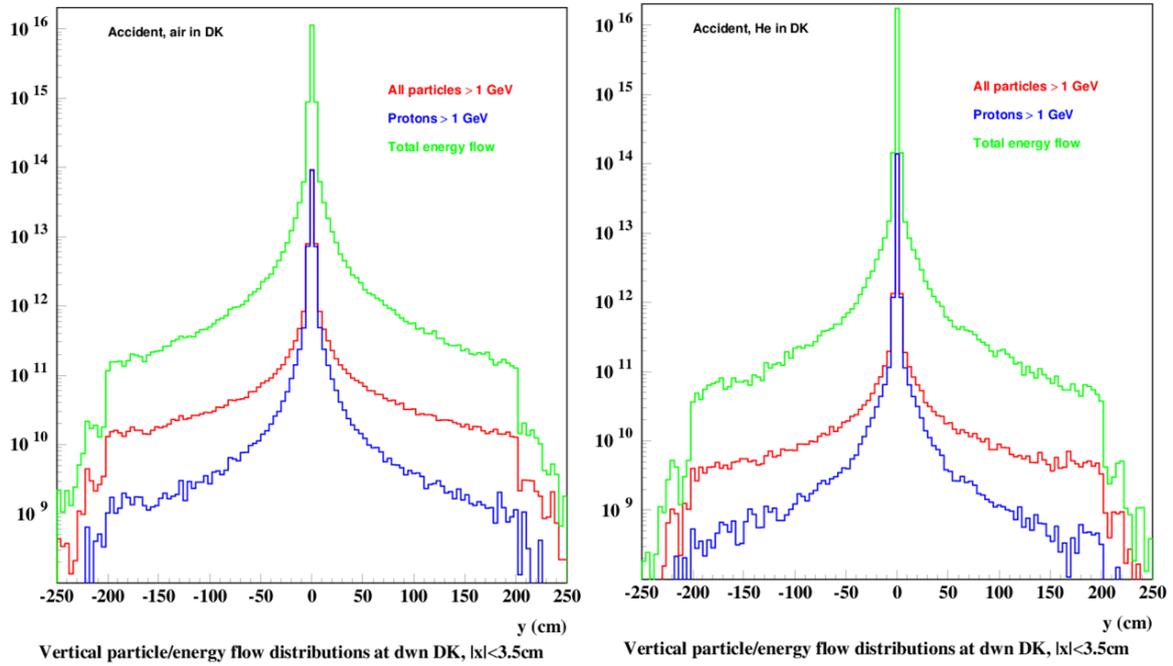

**Figure 6: Gaussian fit (lines) to incoming source calculated with MARS15 code (circles). The data are for the 'target disappears' accident scenario with air (blue) and helium (red) in the decay pipe. Protons with energies above 110 GeV were considered. Normalization is arbitrary**

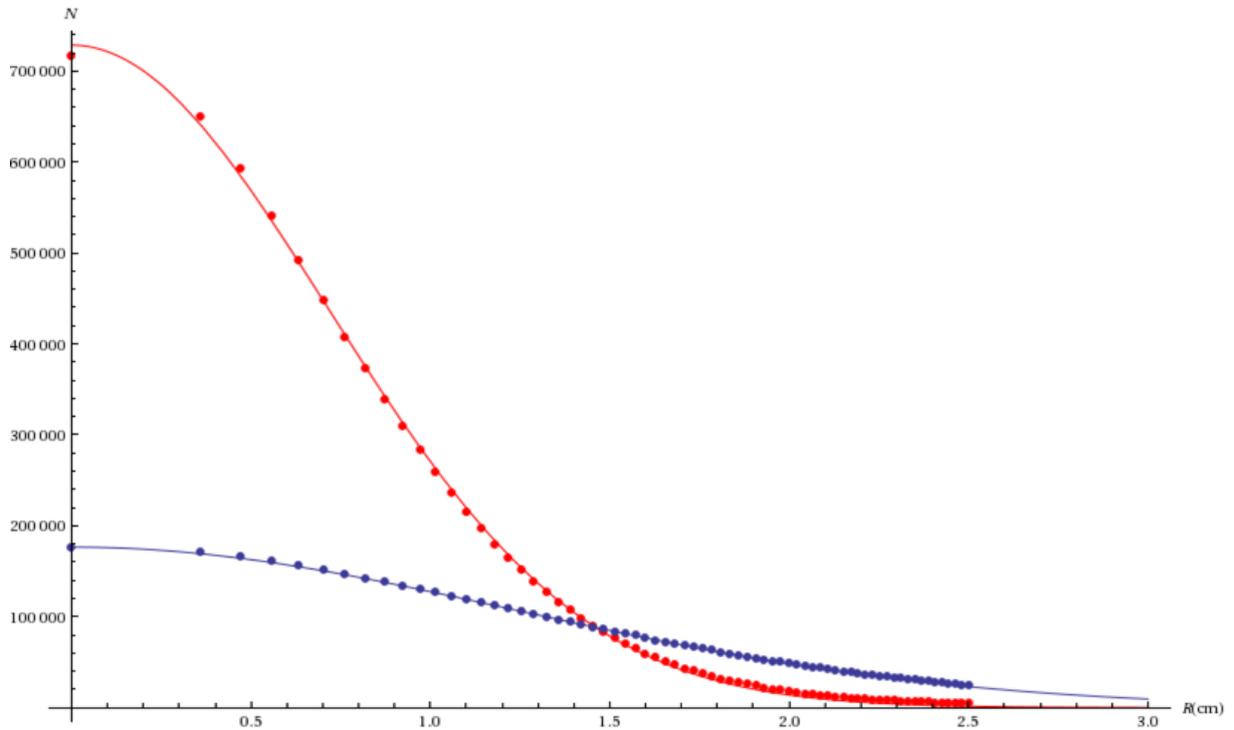

**Table 1: Integral power (kW) deposited in elements of LBNE absorber for the case of normal operation with helium. Neutron energy cutoff is 100 keV**

| Element | Without spoiler | Single Al spoiler | Three Al spoilers |
|---|---|---|---|
| First spoiler |  | 12.6 | 12.5 |
| Aluminium mask | 168.4 | 180.7 | 181.0 |
| Aluminium core | 289.8 | 271.7 | 269.3 |
| Steel core 1 | 24.6 | 16.6 | 19.8 |
| Steel core 2 | 8.6 | 5.7 | 6.9 |
| Steel core 3 | 3.5 | 2.4 |  |
| Steel core 4 | 1.5 | 1.1 |  |
| Steel shielding | 238.3 | 235.5 | 243.3 |
| Total | 734.6 | 726.2 | 732.7 |

**Table 2: Integral power (kW) deposited in aluminium core of LBNE absorber for the case of normal operation with helium. Neutron energy cutoff is 100 keV**

| Aluminium block number | Without spoiler | Single Al spoiler | Three Al spoilers |
|---|---|---|---|
| 1 | 26.9 | 42.8 | 42.5 |
| 2 | 45.0 | 53.7 | 23.9 |
| 3 | 54.2 | 50.8 | 20.6 |
| 4 | 50.7 | 41.6 | 42.0 |
| 5 | 41.5 | 31.7 | 15.1 |
| 6 | 31.7 | 23.2 | 36.3 |
| 7 | 23.3 | 16.5 | 32.0 |
| 8 | 16.7 | 11.6 | 25.0 |
| 9 |  |  | 18.6 |
| 10 |  |  | 13.4 |
| Total | 289.8 | 271.7 | 269.3 |

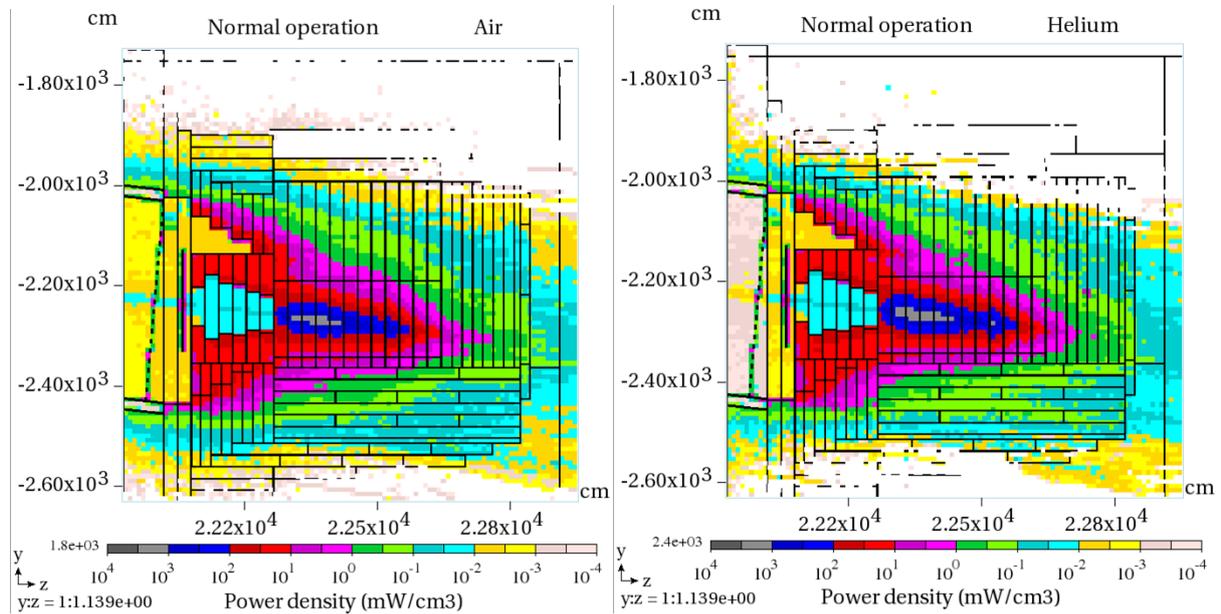

**Figure 7: Calculated power density distributions for normal operation with air (left) and helium (right) in the decay pipe**

As a result of the Monte Carlo modeling, very detailed power density distributions for the hottest regions in the absorber core are provided for the subsequent thermal and stress analysis with the ANSYS code. At present, several options are under investigations that have the potential of providing reduced peak power density in the case of helium compared to the initial estimate of 2.4 mW/cm3. According to extensive ANSYS studies performed for this case, one can expect the normal operation with helium at temperature in the aluminum core not exceeding 100º C.

**Acknowledgements**

This work was supported by Fermi Research Alliance, LLC, under contract No. DE-AC02-07CH11359 with the U.S. Department of Energy.